\documentclass{elsart}

\usepackage{graphicx}
\usepackage{amssymb}

\begin{document}

\begin{frontmatter}
\title{Measurements of Primary and Atmospheric Cosmic-Ray Spectra 
  with the BESS-TeV Spectrometer}
\author[tokyo]{S.~Haino\corauthref{cor1}},
\corauth[cor1]{Corresponding author.}
\ead{haino@icepp.s.u-tokyo.ac.jp}
\author[tokyo]{T.~Sanuki},
\author[kobe] {K.~Abe},
\author[tokyo]{K.~Anraku\thanksref{kngw}},
\author[tokyo]{Y.~Asaoka\thanksref{icrr}},
\author[kek]  {H.~Fuke},
\author[tokyo]{M.~Imori},
\author[kobe] {A.~Itasaki},
\author[kobe] {T.~Maeno\thanksref{cern}},
\author[kek]  {Y.~Makida},
\author[tokyo]{S.~Matsuda},
\author[tokyo]{N.~Matsui},
\author[tokyo]{H.~Matsumoto},
\author[gsfc] {J.W.~Mitchell},
\author[gsfc] {A.A.~Moiseev},
\author[tokyo]{J.~Nishimura},
\author[kobe] {M.~Nozaki},
\author[tokyo]{S.~Orito\thanksref{dead}},
\author[gsfc] {J.F.~Ormes},
\author[gsfc] {M.~Sasaki},
\author[umd]  {E.S.~Seo},
\author[kobe] {Y.~Shikaze},
\author[gsfc] {R.E.~Streitmatter},
\author[kek]  {J.~Suzuki},
\author[kobe] {Y.~Takasugi}, 
\author[kek]  {K.~Tanaka},
\author[kobe] {K.~Tanizaki},
\author[isas] {T.~Yamagami}, 
\author[kek]  {A.~Yamamoto},
\author[tokyo]{Y.~Yamamoto},
\author[kobe] {K.~Yamato},
\author[kek]  {T.~Yoshida},
\author[kek]  {K.~Yoshimura}
\address[tokyo]{The University of Tokyo, Bunkyo, Tokyo 113-0033, Japan}
\address[kobe] {Kobe University, Kobe, Hyogo 657-8501, Japan}
\address[kek]  {High Energy Accelerator Research Organization (KEK),
                Tsukuba, Ibaraki 305-0801, Japan}
\address[gsfc] {National Aeronautics and Space Administration (NASA), 
                Goddard Space Flight Center (GSFC), 
		Greenbelt, MD 20771, USA.}
\address[umd]  {University of Maryland, College Park, MD 20742, USA}
\address[isas] {The Institute of Space and Astronautical Science (ISAS) of 
                Japan Aerospace Exploration Agency (JAXA), 
		Sagamihara, Kanagawa, 229-8510, Japan}
\thanks[kngw]{Present address: Kanagawa University,
              Yokohama, Kanagawa 221-8686, Japan}
\thanks[icrr]{Present address: ICRR, The University of Tokyo, 
              Kashiwa, Chiba 227-8582, Japan}
\thanks[cern]{Present address: CERN, CH-1211 Geneva 23, Switzerland}
\thanks[dead]{deceased.}

\begin{abstract}
Primary and atmospheric cosmic-ray spectra were precisely measured with the 
BESS-TeV spectrometer. The spectrometer was upgraded from BESS-98 to achieve 
seven times higher resolution in momentum measurement. We report absolute 
fluxes of primary protons and helium nuclei in the energy 
ranges, 1--540~GeV and 1--250~GeV/n, respectively, and absolute flux 
of atmospheric muons in the momentum range 0.6--400~GeV/$c$. 
\end{abstract}

\end{frontmatter}

\section{Introduction}
\label{sec:introduction}
The absolute flux and spectral shape of primary cosmic rays are the basis 
to discuss the origin and the propagation history of the cosmic rays 
in the Galaxy. 
The spectrum is also essential as an input to calculate spectra of cosmic-ray 
antiprotons and positrons which are secondary products of cosmic-ray 
interactions with the interstellar gas. Recently, 
the importance of the cosmic-ray spectrum has been emphasized in connection 
with the study of neutrino oscillation observed in atmospheric 
neutrinos~\cite{SuperK}. In contrast with a long-baseline experiment, 
such as K2K~\cite{K2Kexp}, 
the feature of the oscillation study with 
atmospheric neutrinos lies in its wide energy range. 
For example, the Super Kamiokande water Cherenkov detector~\cite{SKdetc} 
can observe neutrinos from 0.1 to 100~GeV.
In order to estimate an accurate flux of atmospheric neutrinos up to 
around 100~GeV, we need to know the primary cosmic-ray flux well above 
100~GeV and the feature of hadronic interactions in that energy region. 
It is crucial to measure primary cosmic-ray flux by determining 
the absolute energy of particles. 
Measurements of atmospheric muon spectrum is also important to check and 
improve our understanding of hadronic interactions. 

The Balloon-borne Experiment with a Superconducting Spectrometer 
(BESS) \cite{ASTRMG,BESSd1} 
has been carried out since 1993 to perform highly sensitive searches 
for cosmic-ray antiparticles and precise measurements of absolute 
flux of various cosmic-ray components. 
The absolute fluxes of primary protons and helium nuclei were precisely 
measured in the energy ranges, 1--120~GeV and 1--54~GeV/n, respectively, 
by a balloon 
observation in 1998~\cite{BESS98}. The overall uncertainties in the 
measurements were less than 5~\% for protons and 10~\% for helium nuclei. 
The absolute flux of atmospheric muons was measured 
from 0.6 to 20~GeV/$c$ at sea level~\cite{GESS95} 
and from 0.6 to 100~GeV/$c$ 
at a mountain altitude~\cite{MESS99}. 
The overall uncertainty in the measurements was less than 10~\%. In order 
to extend the energy range of the precise measurements of cosmic-ray flux 
up to higher energy, the BESS-TeV spectrometer was developed. 
A magnetic rigidity ($R\equiv pc/Ze$) of an incident particle was accurately 
measured by a new tracking system. The maximum detectable rigidity (MDR) 
was significantly improved from 200~GV to 1.4~TV. 

Cosmic-ray observations were carried out with the BESS-TeV spectrometer 
at a balloon altitude and at sea level in 2002. 
We report precise measurements of absolute fluxes of 
primary protons and helium nuclei, and atmospheric muons. 

As to primary cosmic rays, we focused on a energy range above 1~GeV/n in 
this paper. The BESS spectrometer has measured primary cosmic-ray flux 
down to 0.2~GeV/n, which provides significant information on the solar 
modulation effect and the propagation of cosmic rays~\cite{JZWang}. 
Measurements of the low-energy cosmic-ray flux with the BESS-TeV spectrometer 
will be reported elsewhere together with a series of BESS 
balloon-flight data~\cite{Shikaz}. 

\section{The BESS-TeV spectrometer}
\label{sec:detector}

The BESS spectrometer~\cite{BESSd2} was upgraded to 
achieve a significantly high rigidity resolution. 
The upgraded spectrometer, ``BESS-TeV''~\cite{BESStv}, was equipped with 
newly-developed drift chambers. 

\begin{figure}[ht]
  \begin{center}
    \includegraphics[width=10.0cm]{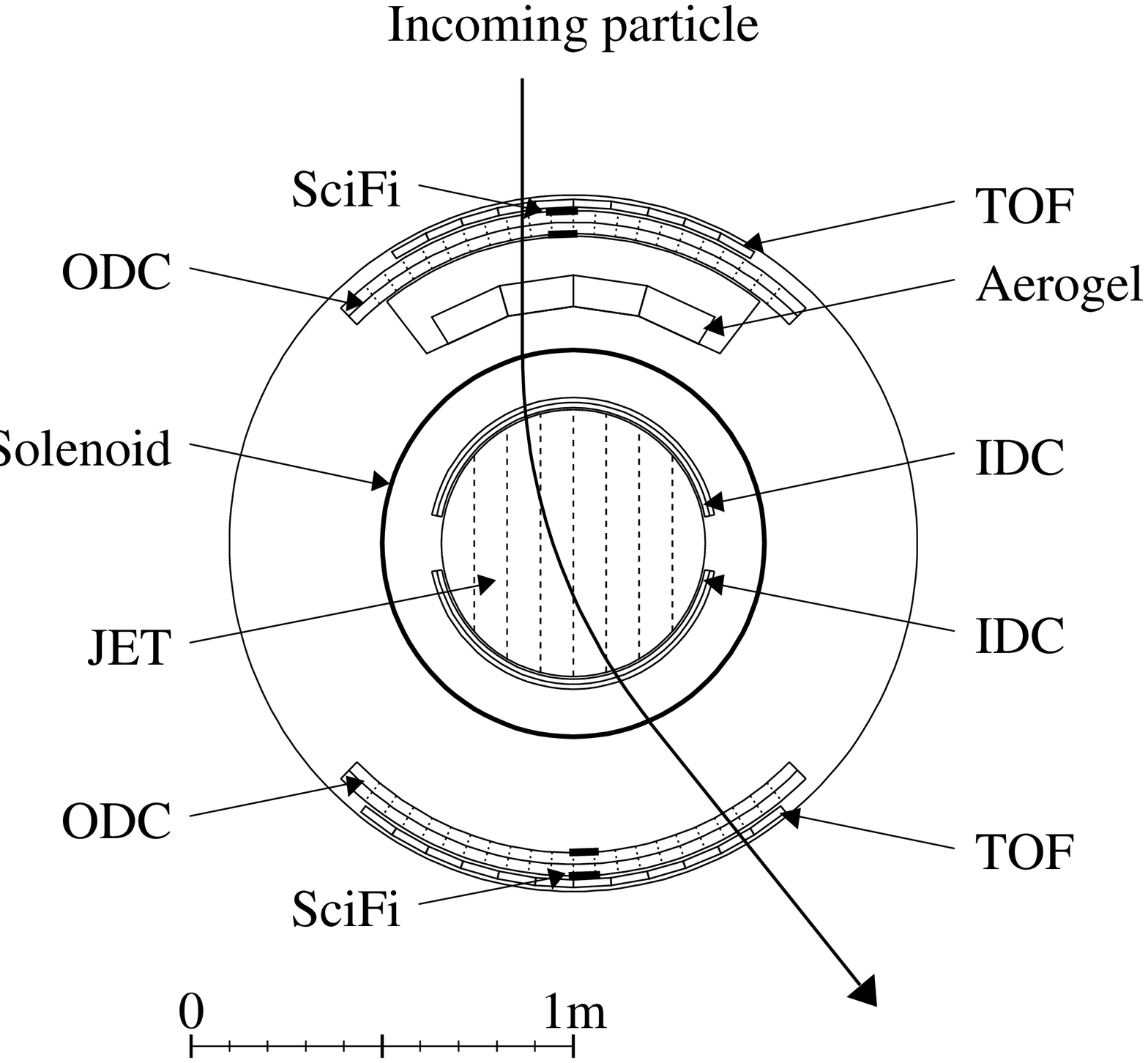}
    \caption{Cross-sectional view of the BESS-TeV spectrometer.}
    \label{fig:detector}
  \end{center}
\end{figure}

As shown in Fig.~\ref{fig:detector}, 
the detector has a unique feature of a cylindrical configuration 
realized by a thin superconducting solenoid~\cite{magnet}. 
The configuration resulted in a large and almost constant geometrical 
acceptance and a uniform detector performance for 
various incident angles and positions. 

\begin{figure}[ht]
  \begin{center}
    \includegraphics[width=12.0cm]{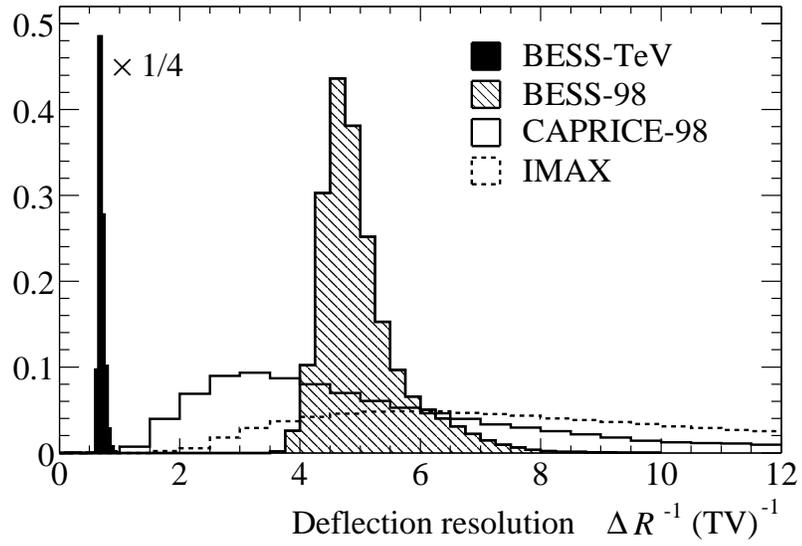}
    \caption{Distribution of the deflection resolution for protons 
      evaluated for each event in the track-fitting procedure. 
      Each area of the histogram is normalized to unity.}
    \label{fig:defreso}
  \end{center}
\end{figure}

In the central region, the solenoid with a diameter of 1~m 
provides a uniform magnetic field of 1~T. The field variation 
is less than 2.5~\% along a typical trajectory of an incoming particle. 
A deflection ($R^{-1}$) of the trajectory is measured by 
a central jet-type drift chamber (JET), 
two inner drift chambers (IDCs) and two outer drift chambers (ODCs), 
all of which were newly developed for the BESS-TeV spectrometer. 
Inside JET and IDCs a trajectory was determined by 
simple circular fitting~\cite{Karimk} using up to 52 hit points. 
Each hit point was measured with a spacial resolution of 150~$\mu$m. 
More accurate deflection was measured by adding 
8 hit points inside ODCs which are placed outside the solenoid 
at a radius of 0.8 m. The hit positions inside the ODCs were used 
above 100~GV, where the effect of multiple scattering in 
the detector material is negligibly small. 
Fig.~\ref{fig:defreso} shows a distribution of the deflection resolution 
($\Delta R^{-1}$) obtained with all the drift chambers. 
The deflection resolution was evaluated in the track-fitting procedure 
for cosmic-ray protons. Those of other spectrometers used 
in previous balloon experiments~\cite{BESS98,CAPP98,IMAX92} are also shown. 
Each area of the histogram is normalized to unity. 
The peak position of 0.7~TV$^{-1}$ corresponds to the MDR of 1.4~TV. In 
order to provide an absolute reference position for their calibration, 
a scintillating fiber counter system (SciFi) was attached to the 
top and the bottom walls of the ODCs. 
SciFi consists of two layers of 1~mm~$\times$~1~mm square-shaped 
scintillating fibers and covers the central cell of each ODC. 
The scintillating fiber layers are shifted by 0.5~mm with each other. 
SciFi can measure a hit position with an accuracy of 
the overlap width of two fiber layers, i.e. 0.5~mm. With a sufficiently large 
number of events, the track position can be 
determined with a better spatial resolution than that of ODCs. 

Time-of-flight (TOF) hodoscopes~\cite{newtof} provide the velocity 
($\beta$) and energy loss (d$E$/d$x$) measurements. 
A $\beta^{-1}$ resolution of 1.4~\% was achieved in the experiment. 
The data acquisition sequence is initiated by a first-level TOF trigger, 
which is a simple coincidence of signals from the upper and lower 
TOF counters. The trigger efficiency was evaluated to be 99.4$\pm$0.2~\% 
by a secondary proton beam at KEK 12~GeV proton synchrotron. The trigger 
rates were 1~kHz and 30~Hz at a balloon altitude and sea level, respectively. 
During the balloon observation, one out of every ten events were recorded 
to sample unbiased trigger events. An auxiliary trigger is generated 
by a signal from a Cherenkov counter~\cite{aerogl} to record 
particles above threshold energy without bias or sampling. The efficiency 
of the Cherenkov trigger was evaluated as the ratio of the number 
of Cherenkov-triggered events to the unbiased triggered events. 
It was found to be 94.1$\pm$~2.0\% and 97.3$\pm$4.0~\% for relativistic 
protons and helium nuclei, respectively. For the determination of 
primary proton and helium fluxes, the Cherenkov-triggered 
events were used above 10~GeV and 5.5~GeV/n, respectively, and 
the TOF-triggered events were used below these energies. 
During the ground observation, all the TOF-triggered events were 
recorded to determine the atmospheric muon flux. 

\section{Observations}
\label{sec:observations}

The BESS-TeV spectrometer was launched by a balloon from 
Lynn Lake, Manitoba, Canada (56.5$^{\circ}$N, 101.0$^{\circ}$W), 
on 7th August 2002. After about four-hour ascending, the payload 
reached a floating altitude of 37 km (residual atmosphere of 4.8~g/cm$^2$). 
The geomagnetic cutoff rigidity was 0.5~GV or smaller throughout the flight. 
The total live time of the data-taking was 38,215 seconds (10.6 hours) 
during the floating period. 
Among them 11~\% (1.2 hours) of data taken around the 
sun rise were not used in this analysis, because the rapid temperature 
variation might introduce a large systematic error in the chamber calibration. 

\begin{figure}[ht]
  \begin{center}
    \includegraphics[width=12.0cm]{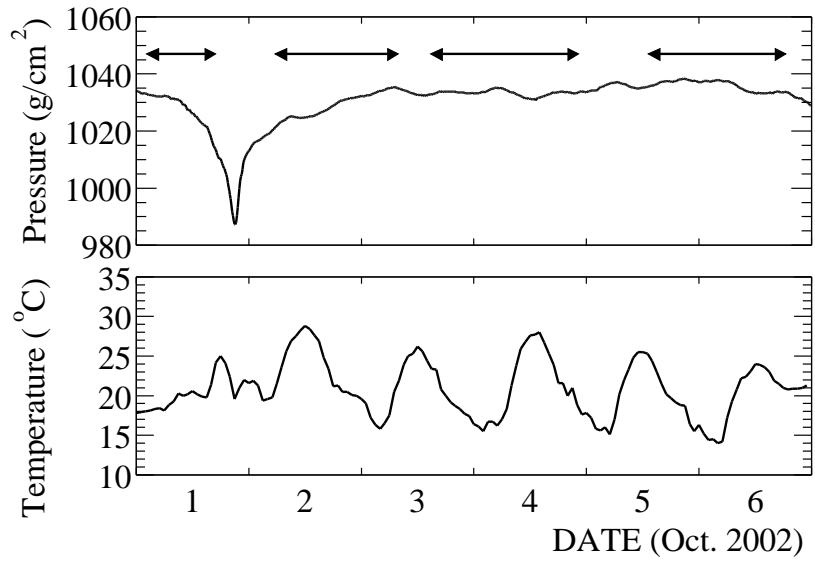}
    \caption{Atmospheric pressure and temperature during the 
      ground observation. Arrows indicate the data-taking periods 
      used to determine the atmospheric muon flux.
      The atmospheric temperature data were obtained 
      from Ref.~\cite{jmatmp}.}
    \label{fig:tsukuba}
  \end{center}
\end{figure}

The ground observation was carried out at KEK located at Tsukuba, Japan 
(36.2$^{\circ}$N, 140.1$^{\circ}$W), during a period of 1st--6th October 
2002. Tsukuba is located 30~m above sea level with the vertical 
cutoff rigidity of 11.4~GV. The variations in the atmospheric pressure 
and temperature during the observation are shown in Fig.~\ref{fig:tsukuba}. 
The atmospheric temperature data were obtained from Ref.~\cite{jmatmp}. 
The arrows indicate the data-taking periods used to determine the 
atmospheric muon flux. We did not use the data during a period where the 
variation in atmospheric pressure was large. The mean (root-mean-square) 
atmospheric pressure and temperature during the periods used for the flux 
determination were 1032.2~g/cm$^2$ (4.4~g/cm$^2$) and 20.8~$^{\circ}$C 
(3.7~$^{\circ}$C), respectively. The total live time of the data-taking 
was 329,403 seconds 
(91.5 hours). 

\section{Data analysis}
\label{sec:analysis}

\subsection{Event reconstruction}
\label{sec:reconstruction}

Each hit position inside the drift chambers was calculated from 
the drift time digitized by a flash analog-to-digital converter. 
The calculation was carried out based on a relation between the hit position 
and the drift time ($x$-$t$ relation). The $x$-$t$ relation was precisely 
calculated by a drift chamber simulation package, GARFIELD~\cite{GARFLD}, 
and a gas property simulation package, MAGBOLTZ~\cite{MAGBLZ}. 
Although the chambers were constructed carefully 
with a tolerance of 100~$\mu$m, there was a small position deviation 
of wires and field-shaping patterns,  which could locally modify the 
electric field. In order to take account of the limited accuracy in 
the chamber manufacturing, a correction was commonly applied to 
the calculated $x$-$t$ relation throughout the experiments. 
The correction was obtained to minimize the $\chi^2$ in the fitting of 
straight tracks of clean muon events observed on the ground without 
magnetic field. 
The correction was as small as expected from the accuracy in the 
chamber manufacturing. 
During the observations, 
the $x$-$t$ relation was affected by the variation in the pressure and 
temperature of the chamber gas. In order to take account of 
these time-dependent variations, the $x$-$t$ relation was calibrated 
for each data-taking run. 
Especially in calibrating the $x$-$t$ relation of ODCs,  
an absolute reference positions were provided by SciFi, 
which are not affected by the variation in the pressure nor temperature. 

The same calibration procedure was applied to the muon events observed on 
the ground without magnetic field, where the straight tracks were used as 
an absolute reference of events with infinite rigidity. 
We checked the deflection of the muon events by applying 
circular-fitting to the calibrated hit points inside the JET and IDCs. 
The mean value of the obtained deflection was smaller than (10~TV)$^{-1}$. 
Therefore, the systematic shift in the deflection measurement originated 
from this calibration procedure should be smaller than (10~TV)$^{-1}$. 

The precise alignment of the ODCs with respect to the JET was determined 
by checking the consistency between tracks 
reconstructed by JET and hit points measured by ODCs. 
In order to minimize the effect of multiple scattering 
in the detector material, we selected events whose rigidity was 
measured to be higher than 4~GV by JET and IDCs. 
The chamber alignment was calibrated for each run. We checked the 
variation in the calibrated position of ODCs in 84 runs during the 
ground observation. The root-mean-square of the variation was smaller 
than 20~$\mu$m. This variation might introduce a systematic error in 
the deflection measurement of $\Delta R^{-1}$ = (5.3~TV)$^{-1}$.

\subsection{Event selection}
\label{sec:selection}

We selected events with a single track fully contained 
inside the fiducial volume defined 
by the central six columns out of eight columns in the JET. 
It ensured that the track is long enough for reliable 
rigidity measurement. In order to obtain a nearly vertical flux 
of atmospheric muons, an additional 
cut on the zenith angle ($\theta_{\rm z}$) was applied as 
$\cos\theta_{\rm z} > 0.98$ below 20~GV and 
$\cos\theta_{\rm z} > 0.90$ above 20~GV. 
We have checked the change of the atmospheric muon flux above 20~GV 
by requiring two conditions, 
$\cos\theta_{\rm z} > 0.98$ and $\cos\theta_{\rm z} > 0.90$. 
We found that the requirement to zenith angle could be relaxed 
to $\cos\theta_{\rm z} > 0.90$ without changing the observed flux. 
For protons and helium nuclei zenith angle cut was not applied. 
Due to the detector geometry, however, zenith angle is limited as 
$\cos\theta_{\rm z} > 0.80$ above 2~GV. 

A single-track event was defined as an 
event which has only one isolated track inside the JET and one or two 
hit counters in each layer of the TOF hodoscopes. The single-track 
selection eliminated rare interacting events. 
To estimate the efficiency of the single-track selection, 
Monte Carlo simulations with GEANT3/4~\cite{GEANT3,GEANT4} were performed. 
The probability that each particle could pass through the selection was 
evaluated by applying the same selection criteria to the Monte Carlo events. 
The resultant efficiency of the single-track event selection 
for protons was 82.0$\pm$2.6~\% at 20~GV and 79.9$\pm$4.0~\% at 200~GV, 
the efficiency for helium nuclei was 69.6$\pm$6.6~\% at 20~GV 
and 66.9$\pm$7.2~\% at 200~GV, and 
the efficiency for muons was 96.8$\pm$1.5~\% at 20~GV 
and 95.9$\pm$1.5~\% at 200~GV. 
In order to ensure the track reconstruction in the $z$-direction 
(perpendicular to the bending plane), we required a consistency of 
hit position in the $z$-direction measured by 
two independent detectors; the drift chamber and the TOF hodoscope. 
The efficiency of this selection was estimated 
to be 97.2$\pm$1.0~\% for protons, 95.5$\pm$1.5~\% for helium nuclei, 
and 98.1$\pm$1.0~\% for muons. 

\begin{figure}[ht]
  \begin{center}
    \includegraphics[width=12.0cm]{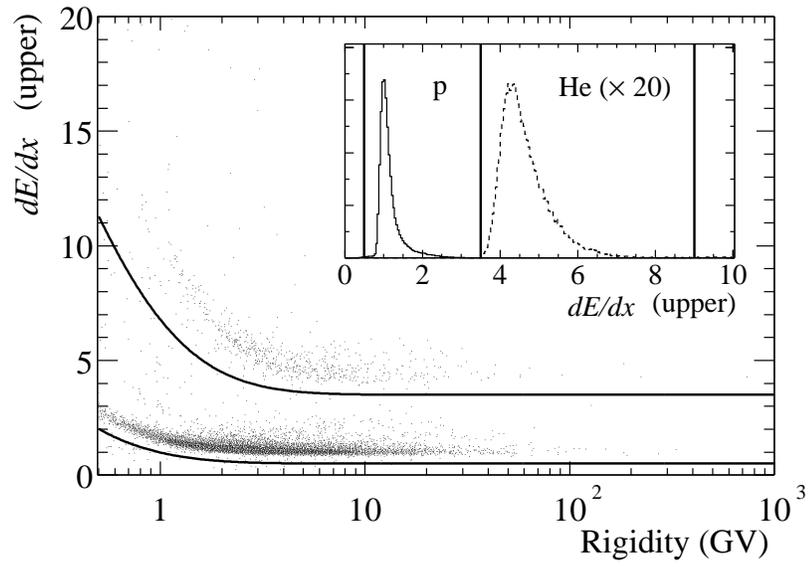}
    \caption{Proton band in d$E$/d$x$ (top TOF) vs. rigidity 
      obtained from the balloon observation. A d$E$/d$x$ in the bottom 
      TOF was also checked. The superimposed graph shows the selection 
      criteria for protons and helium nuclei above 10~GV.}
    \label{fig:dedx}
  \end{center}
\end{figure}

\begin{figure}[ht]
  \begin{center}
    \includegraphics[width=12.0cm]{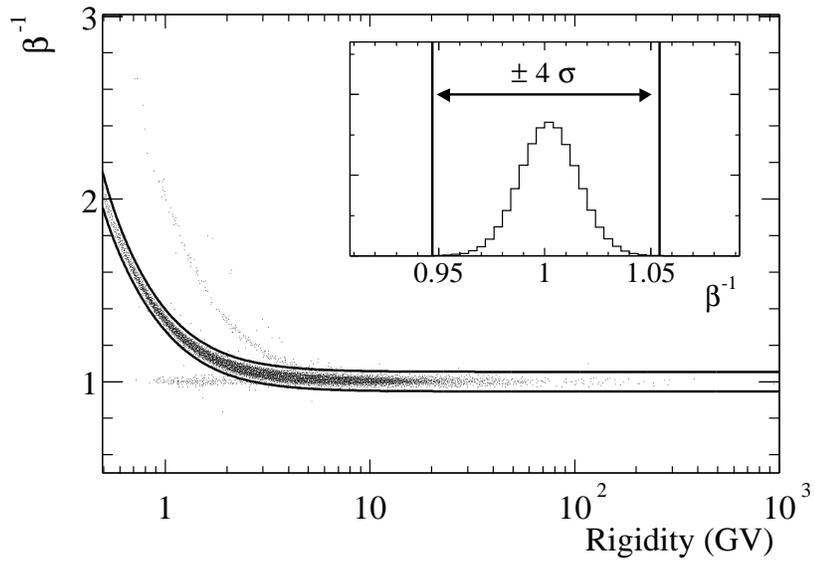}
    \caption{Scatter plot of $\beta^{-1}$ vs. rigidity 
      obtained from the balloon observation after proton d$E$/d$x$ selection. 
      The superimposed graph shows the selection criteria for 
      protons above 10~GV.}
    \label{fig:beta}
  \end{center}
\end{figure}

Particle identification was performed by requiring proper d$E$/d$x$ 
measurements with both upper and lower layers of the TOF hodoscopes and 
$\beta^{-1}$ as functions of rigidity. 
Figs.~\ref{fig:dedx} and \ref{fig:beta} show the selection criteria for 
protons. Helium nuclei and 
muons were identified in the same manner. The efficiencies of d$E$/d$x$ 
selection were estimated with another sample selected by independent 
measurement of energy loss inside the JET. We found that 96.3$\pm$0.3~\% 
of protons, 92.3$\pm$0.9~\% of helium nuclei and 99.9$\pm$0.1~\% of 
muons were properly identified. 
Since the $\beta^{-1}$ distribution is well described by Gaussian and a 
half-width of the $\beta^{-1}$ selection band was set at 4 $\sigma$, 
the efficiency is very close to unity. 
Following this particle identification procedure, 444,578 proton, 
38,006 helium, and 688,983 muon candidates were obtained. 

\subsection{Backgrounds}
\label{sec:backgrounds}

As shown in Fig.~\ref{fig:beta}, protons were identified without 
contamination by examining $\beta^{-1}$ distribution below 2~GV. 
Above 2~GV, there were contamination of light particles such as 
positrons, pions and muons. 
The ratio of light particles to protons was observed to be 
4.5~\% at 1~GV and 1.1~\% at 2~GV, which was expressed by a 
power law with an index of $-2.1$. This tendency was reproduced 
well by a Monte Carlo simulation~\cite{mhonda} based on the DPMJET-III event 
generator~\cite{dpmjet}. Above 10~GV, the simulated $\mu^{+}$/p ratio 
shows almost constant value of 0.2~\%. In this analysis the contamination of 
the light particles were subtracted by assuming the power law, which 
would underestimate the background of light particles. 
However, The contamination is smaller than the statistical errors. 
Above 3~GV, there were contamination of deuterons, which was 
observed to be 2~\% at 3~GV. 
According to our previous measurement~\cite{JZWang}, the d/p flux ratio 
decreases with increasing rigidity. The ratio should follow a 
decrease in escape path lengths of primary cosmic-ray nuclei~\cite{Englmn}. 
Since the contamination of deuterons was as small as the statistical errors, 
and should decrease with increasing rigidity, 
no subtraction was made for the deuteron contamination. 
Therefore, above 3~GV hydrogen nuclei were selected, which included 
a small amount of deuterons. 
Helium nuclei were identified clearly by redundant charge measurement 
inside the upper and lower TOF hodoscopes. 
The helium candidates contained both $^3$He and $^4$He. 
In conformity with previous experiments, all doubly charged particles 
were treated as $^4$He. 

Among the muon candidates observed on the ground, there were 
contaminations of electrons, positrons and protons. 
In our previous work, the ratio of electrons and positrons 
to muons was measured to be 1.5~\% at 0.5~GV and 0.3~\% at 1~GV 
with an electro-magnetic shower counter~\cite{GESS95}. 
Since the BESS-TeV spectrometer is not equipped with an electro-magnetic 
shower counter, no subtraction was made for the contamination of 
electrons and positrons. However, the contamination was smaller than 
the statistical errors. 
Protons were identified by examining $\beta^{-1}$ distribution below 2.5~GV.
The p/$\mu$ flux ratio was observed to be 3.3~\% at 1~GV and 1.2~\% at 2.5~GV, 
which was expressed by a power law with an index of $-1.0$. 
Below 7~GV, the power law agreed well with a previous measurement 
of the p/$\mu$ ratio~\cite{Durham}, which shows, however, 
constant value of about 0.4~\% above 7~GV. 
The proton background was subtracted from muon candidates 
assuming the power law below 7~GV and constant value of 0.4~\% above 7~GV. 

\subsection{Normalization and corrections}
\label{sec:normalization}

In order to obtain the absolute flux of protons, helium nuclei and muons 
at the top of the detector, energy loss by ionization inside the detector 
material, live time and geometrical acceptance were estimated. 
The energy of each incoming particle was calculated by integrating the 
energy losses inside the detector tracing back along the particle trajectory. 
The total live time of the data-taking was measured exactly 
by counting 1 MHz clock pulses with a scaler system 
gated by a ``ready'' status that controls the first-level trigger. 
The geometrical acceptance defined for this analysis was calculated 
as a function of rigidity with a simulation technique~\cite{Sulvin}. 
The geometrical acceptance is 0.0886$\pm$0.0003 m$^2$sr for 
protons and helium nuclei at 10~GV and 0.0302$\pm$0.0001 m$^2$sr for 
muons at 10~GV. The acceptance for muons is about 1/3 of that for protons 
and helium nuclei because of the additional requirement on the zenith 
angle ($\cos\theta_{\rm z} > 0.98$) as described in 
Section~\ref{sec:selection}. 
The simple cylindrical shape and the uniform 
magnetic field make it simple and reliable to determine the 
geometrical acceptance precisely. The error on the acceptance calculation  
was estimated to be 0.3~\% which represents the uncertainty 
of the detector alignment 

In order to obtain the absolute flux of primary protons and helium nuclei 
at the top of the atmosphere, interaction loss and secondary particle 
production in the residual atmosphere were estimated. 
According to the Monte Carlo studies, probabilities for primary cosmic rays 
to penetrate the residual atmosphere of 4.8~g/cm$^2$ are 93.8$\pm$0.7~\% 
and 91.3$\pm$2.0~\% for protons and helium nuclei, respectively, at 10~GeV 
and almost constant over the entire rigidity range discussed here. 
Atmospheric secondary protons were subtracted 
based on a calculation for the maximum solar activity epoch by 
Papini et al.~\cite{Papini}. A secondary-to-primary proton ratio is 
4.8$\pm$0.5~\% at 1~GeV and 1.7$\pm$0.2~\% above 10~GeV. Atmospheric 
secondary helium nuclei above 1~GeV/n are dominated by the fragments 
of heavier cosmic-ray nuclei (mainly carbon and oxygen). 
The flux ratio of the atmospheric secondary helium to the primary carbon 
and oxygen was estimated to be 14~\% at a depth of 4.8 g/cm$^2$, 
based on the total inelastic cross sections of CNO+Air interactions 
and the helium multiplicity in $^{12}$C + CNO interactions~\cite{AhmadK}. 
The total correction of atmospheric 
secondary helium produced by all the primary nuclei with Z$>$2 was 
estimated to be 1.6$\pm$0.5~\% at 1~GeV/n and 2.1$\pm$0.6~\% at 10~GeV/n.

\subsection{Spectrum deformation effect}
\label{sec:syserror}

\begin{figure}[ht]
  \begin{center}
    \includegraphics[width=12.0cm]{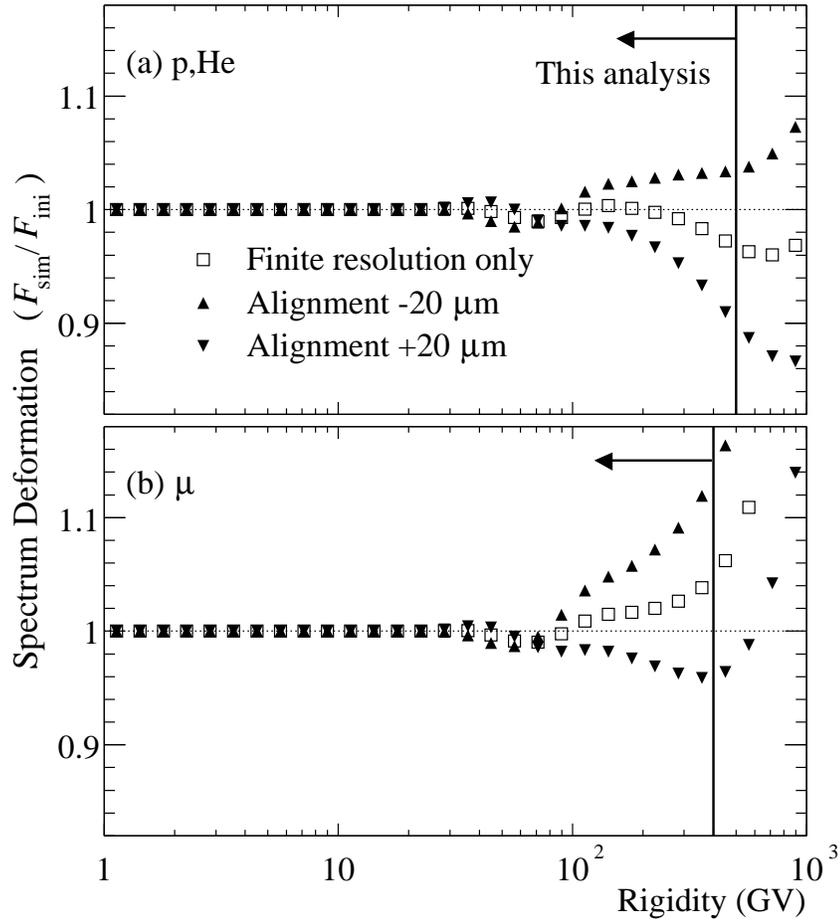}
    \caption{Spectrum deformation effects for 
      (a) primary protons and helium nuclei and (b) atmospheric muons. 
      The deformation effects are shown 
      as ratios of simulated spectra ($F_{\rm sim}$) to the input 
      spectra ($F_{\rm ini}$) as functions of rigidity. Open squares 
      show the case when ODCs are aligned correctly. Upward and downward 
      triangles show the case when ODCs are artificially displaced by 
      $\pm20 \mu$m. }
    \label{fig:deform}
  \end{center}
\end{figure}

Because of the limited accuracy of the rigidity measurement and the steep 
spectral shape, the observed spectrum may suffer deformation. 
The limited accuracy of the rigidity measurement was decomposed 
into two sources: 
(i)~a finite resolution in the rigidity measurement, and 
(ii)~a small shift in the measured deflection 
which is caused by the misalignment of the ODCs. 
We estimated the systematic error in the spectrum measurement 
caused by such limited accuracy of the rigidity measurement. 
The systematic error was estimated by a Monte Carlo simulation. 
In the simulation, detailed response of the drift chamber was implemented, 
such as the distribution of primary ionization clusters, 
the diffusion in the gas, the fluctuation of avalanche gain, 
and the digitization in the readout electronics. 
The simulated rigidity resolution reproduced well the experimental result 
shown in Fig.~\ref{fig:defreso}. Therefore the effect of spectrum deformation 
should be correctly estimated by this simulation. 
As an input spectrum, a power-law function was used with a 
spectral index of $-2.7$ for both primary protons 
and helium nuclei. For atmospheric muons, the spectral index was 
chosen as $-3.2$, which was obtained by fitting a power-law function to the 
muon spectrum obtained with this experiment in the rigidity range 100--400~GV. 
Fig.~\ref{fig:deform} shows the effect of spectrum deformation as a ratio 
of simulated spectrum to the input spectrum. The open squares show the 
spectrum deformation when ODCs are aligned correctly. 
In this case, the deformation is caused only by the effect of (i). 
The change of the spectrum is less than 5~\% for 
protons and helium nuclei below 1~TV, and for muons below 400~GV. 
The upward and downward triangles show the spectrum deformation 
when ODCs are artificially displaced by $\pm20 \mu$m. 
We found that the shift of the measured deflection was smaller 
than (5.3~TV)$^{-1}$ as described in Section~\ref{sec:reconstruction}. 
The change of the spectrum was less than 2~\% below 100~GV 
and 10~\% at 500~GV for primary protons and helium nuclei, 
and less than 3~\% below 100~GV and 13~\% at 400~GV for muons. 
Since the input spectrum of atmospheric 
muons is steeper than that of primary protons and helium nuclei, 
the spectrum deformation effect is more significant for atmospheric muons.
We did not apply any corrections, such as unfolding, on the observed 
spectra. The spectrum deformation effect estimated above was treated 
as a systematic error on the observed spectra, which was as small as 
the statistical error over the entire rigidity range in this analysis. 

\section{Results and discussions}
\label{sec:results}

We have obtained the absolute fluxes of primary protons in the range 
1--540~GeV and helium nuclei in the range 1--250~GeV/n at the top of 
the atmosphere from the BESS-TeV balloon-flight data in 2002. 
We have obtained the absolute flux of muons in the range 0.6--400~GeV/$c$ 
at sea level (30~m a.s.l.) from the BESS-TeV ground-observation data in 2002. 
The results of protons and helium nuclei are summarized in 
Tables~\ref{tab:pflux} and \ref{tab:heflux}, respectively, and 
the results of atmospheric muons are summarized in Table~\ref{tab:muflux}. 
The overall uncertainties including both statistical and systematic errors 
were less than $\pm$15~\% for protons, $\pm$20~\% for helium nuclei, 
and $\pm$20~\% for muons. 

\begin{figure}[ht]
  \begin{center}
    \includegraphics[width=13.5cm]{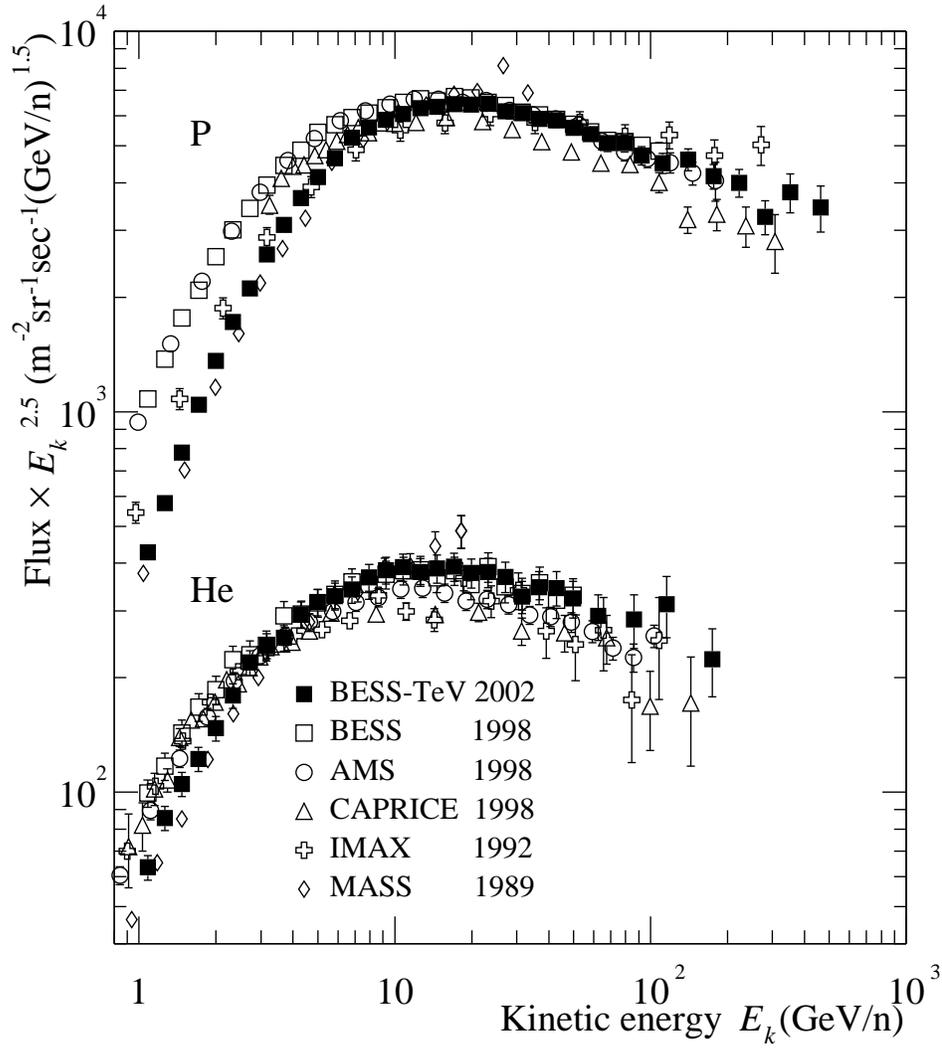}
    \caption{Absolute differential energy spectra of 
      primary protons and helium nuclei multiplied by $E_k^{2.5}$. 
      The spectra obtained by 
      other experiments~\cite{BESS98,CAPP98,IMAX92,AMS-p2,AMS-he,MASP89}
      are also shown.}
    \label{fig:pheflux}
  \end{center}
\end{figure}

The results of primary proton and helium spectra are shown 
in Fig.~\ref{fig:pheflux} in comparison with other experiments with 
magnetic spectrometers~\cite{BESS98,CAPP98,IMAX92,AMS-p2,AMS-he,MASP89}. 
Discrepancies in the observed spectra below 10~GeV 
for protons and 5~GeV/n for helium nuclei come from the difference of 
solar activity (around minimum in 1998 and around maximum in 2002). 
The detailed analysis of the solar modulation affecting the low-energy 
spectra in a series of BESS flights will be discussed elsewhere~\cite{Shikaz}. 

Although a small systematic shift of 2--3~\% was found in absolute 
proton flux between the results of BESS-TeV and BESS-98 from 30 to 100~GeV, 
the results are well consistent within the overall uncertainty of 5~\%. 
The main source of this uncertainty was the aerogel trigger efficiency. 
The accuracy of the efficiency estimation was limited by the statistics 
of the event sample to be 2.0~\% and 3.0~\% for BESS-TeV and BESS-98, 
respectively. 

Our resultant spectral shape of protons and helium nuclei is very similar 
to that measured by AMS~\cite{AMS-p2,AMS-he}. 
Above 30~GeV, the absolute proton flux measured 
by BESS-TeV and AMS shows good agreement within 5~\%. 
However, there is 15~\% discrepancy in the absolute flux of helium nuclei. 
Both proton and helium spectra 
by CAPRICE-98~\cite{CAPP98} show steeper spectra than our results. 

At high energies, the spectrum $F$ may be parameterized by a power 
law in kinetic energy, $E_{\rm k}$, as $F=\Phi E_{\rm k}^{-\gamma}$. 
The fitting range was chosen to be 30--540~GeV for protons 
and 20--250~GeV/n for helium nuclei 
so that the solar modulation effect was negligible. 
The best fit values and uncertainties for protons 
($\Phi_{\rm p}$ and $\gamma_{\rm p}$) and helium nuclei 
($\Phi_{\rm He}$ and $\gamma_{\rm He}$) were obtained as
\[
\Phi_{\rm p} = (1.37 \pm 0.06({\rm sta.}) \pm 0.11({\rm sys.})) 
               \times 10^4 ~(\rm m^2\cdot sr\cdot s\cdot GeV)^{-1}
\]
\[
\gamma_{\rm p} = 2.732 \pm 0.011({\rm sta.}) \pm 0.019({\rm sys.})
\]
and
\[
\Phi_{\rm He} = (7.06 \pm 0.94({\rm sta.}) \pm 1.17({\rm sys.})) 
                \times 10^3 ~(\rm m^2\cdot sr\cdot s\cdot (GeV/n))^{-1}
\]
\[
\gamma_{\rm He} = 2.699 \pm 0.040({\rm sta.}) \pm 0.044({\rm sys.})
\]
respectively. 
The two parameters were strongly correlated, with a correlation 
coefficient of $-0.98$. 

\begin{figure}[ht]
  \begin{center}
    \includegraphics[width=13.5cm]{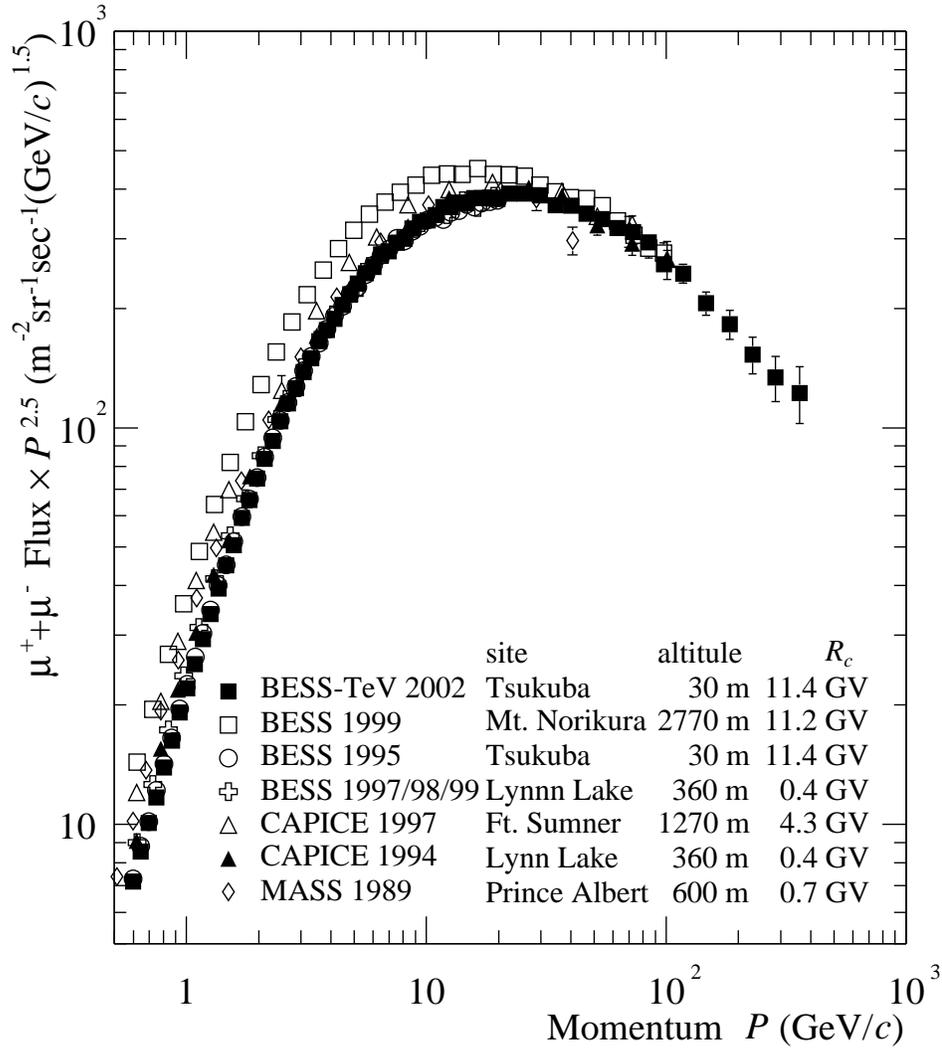}
    \caption{Absolute differential momentum spectrum of atmospheric muons 
      multiplied by $P^{2.5}$. The spectra obtained by other 
      experiments~\cite{GESS95,MESS99,CAPM97,MASM89} are also shown. 
      The geomagnetic cutoff rigidity at each site is indicated as $R_{c}$. 
      The difference below 30~GeV/$c$ among the spectra is mainly 
      due to the different altitudes.}
    \label{fig:muflux}
  \end{center}
\end{figure}

\begin{figure}[ht]
  \begin{center}
    \includegraphics[width=13.5cm]{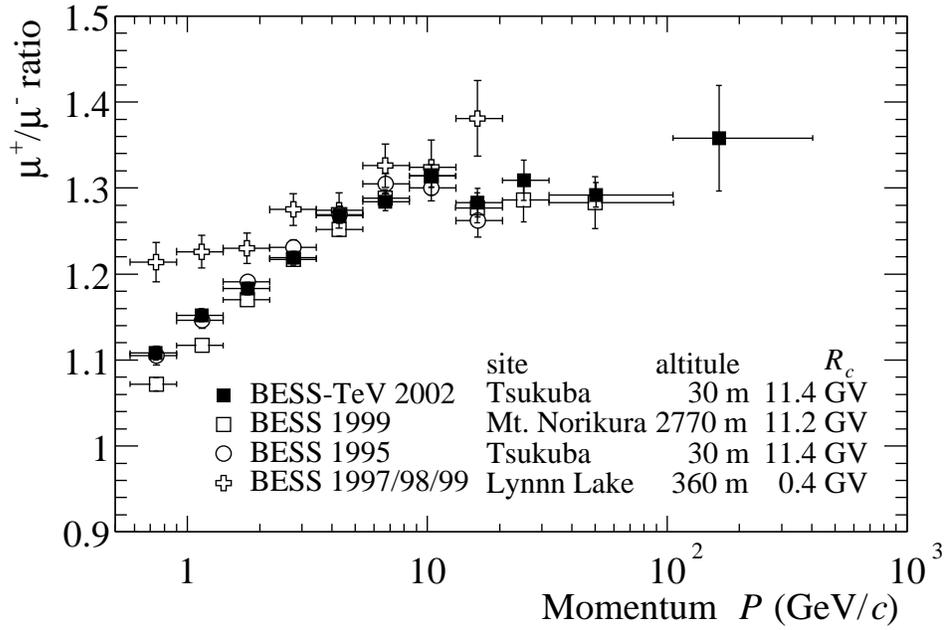}
    \caption{Charge ratios of atmospheric muons. 
      Only statistical errors are included. The charge ratios obtained by 
      other experiments~\cite{GESS95,MESS99} are also shown. 
      The geomagnetic cutoff rigidity at each site is indicated as $R_{c}$. 
      The difference among the charge ratios is mainly due to the difference 
      in geomagnetic cutoff rigidity.}
    \label{fig:muratio}
  \end{center}
\end{figure}

The result of atmospheric muon spectrum is shown in Fig.~\ref{fig:muflux} 
in comparison with other absolute flux measurements by magnetic 
spectrometers~\cite{GESS95,MESS99,CAPM97,MASM89}. 
The discrepancy below 30~GeV/$c$ among the observed 
muon spectra is mainly due to the difference in altitudes. 
Fig.~\ref{fig:muratio} shows charge ratios of atmospheric muons 
observed in a series of muon measurement with the BESS experiments. 
The difference in the charge ratios observed in Japan and Canada 
comes from the different geomagnetic cutoff rigidity~\cite{GESS95}. 

\section{Conclusion}
\label{sec:conclusion}

We have measured energy spectra of primary protons in the range 
1--540~GeV and helium nuclei in the range 1--250~GeV/n by a balloon 
observation, and momentum spectrum of atmospheric muons in the range 
0.6--400~GeV/$c$ by a ground observation at sea level. 

The overall uncertainties were less than 15~\% for protons, less than 20~\% 
for helium nuclei, and less then 20~\% for muons. 
Primary cosmic-ray spectra provide fundamental information 
on the origin and propagation history of the cosmic rays. 
The results also provide accurate input spectra to predict an 
atmospheric neutrino flux. The result of atmospheric muon spectrum 
will improve the accuracy of atmospheric neutrino 
calculation in the wide momentum range. 

\begin{ack}
The authors thank NASA and the National Scientific Balloon Facility 
for their professional and skillful work in carrying out the BESS flight. 
They deeply thank ISAS and KEK for their continuous support 
and encouragement for the BESS experiment. 
They would especially like to thank Doctor T. Taniguchi for his kind 
advice and cooperation to develop the drift chambers. 
This experiment was supported by Grant-in-Aid for Scientific Research 
(12047206 and 12047227) from the Ministry of Education, Culture, Sport, 
Science and Technology, Japan, and was supported by NASA, USA. 
\end{ack}

\clearpage

\clearpage

\clearpage

\clearpage

\clearpage

\clearpage

\clearpage

\clearpage

\clearpage

\clearpage

\renewcommand{\arraystretch}{1.1}

\begin{table}
  \caption{Primary proton flux at the top of the atmosphere.}
  \scriptsize
  \label{tab:pflux}
  \vspace{0.5cm}
  \begin{center}
    \begin{tabular}{rlrl} \hline \hline
      \multicolumn{2}{c}{
	\begin{tabular}{@{}c@{}}       
	  Energy range\\ (GeV)
	\end{tabular}
      }
      & 
      \begin{tabular}{@{}c@{}}       
        $\overline{E_{\rm k}}$\\ (GeV)
      \end{tabular}
      &
      \begin{tabular}{@{}c@{}}       
	Flux$\pm\Delta F_{\rm sta}\pm\Delta F_{\rm sys}$\\
	(m$^{-2}$sr$^{-1}$s$^{-1}$GeV$^{-1}$)
      \end{tabular}
      \\
      \hline
 1.00-- & 1.17 &  1.08 & 3.50 $\pm$0.03 $\pm$0.11 $\times 10^{2}$\\
 1.17-- & 1.36 &  1.26 & 3.22 $\pm$0.03 $\pm$0.10 $\times 10^{2}$\\
 1.36-- & 1.58 &  1.47 & 2.98 $\pm$0.02 $\pm$0.09 $\times 10^{2}$\\
 1.58-- & 1.85 &  1.71 & 2.71 $\pm$0.02 $\pm$0.08 $\times 10^{2}$\\
 1.85-- & 2.15 &  2.00 & 2.41 $\pm$0.02 $\pm$0.07 $\times 10^{2}$\\
 2.15-- & 2.51 &  2.33 & 2.08 $\pm$0.02 $\pm$0.06 $\times 10^{2}$\\
 2.51-- & 2.93 &  2.71 & 1.74 $\pm$0.01 $\pm$0.05 $\times 10^{2}$\\
 2.93-- & 3.42 &  3.16 & 1.45 $\pm$0.01 $\pm$0.04 $\times 10^{2}$\\
 3.42-- & 3.98 &  3.69 & 1.19 $\pm$0.01 $\pm$0.03 $\times 10^{2}$\\
 3.98-- & 4.64 &  4.30 & 9.52 $\pm$0.08 $\pm$0.27 $\times 10^{1}$\\
 4.64-- & 5.41 &  5.01 & 7.35 $\pm$0.07 $\pm$0.21 $\times 10^{1}$\\
 5.41-- & 6.31 &  5.84 & 5.63 $\pm$0.05 $\pm$0.16 $\times 10^{1}$\\
 6.31-- & 7.36 &  6.81 & 4.34 $\pm$0.04 $\pm$0.12 $\times 10^{1}$\\
 7.36-- & 8.58 &  7.93 & 3.15 $\pm$0.03 $\pm$0.09 $\times 10^{1}$\\
 8.58-- & 10.0 &  9.25 & 2.25 $\pm$0.03 $\pm$0.06 $\times 10^{1}$\\
 10.0-- & 11.7 &  10.8 & 1.59 $\pm$0.01 $\pm$0.05 $\times 10^{1}$\\
 11.7-- & 13.6 &  12.6 & 1.12 $\pm$0.01 $\pm$0.04 $\times 10^{1}$\\
 13.6-- & 15.8 &  14.7 & 7.71 $\pm$0.04 $\pm$0.28 \\
 15.8-- & 18.5 &  17.1 & 5.33 $\pm$0.03 $\pm$0.20 \\
 18.5-- & 21.5 &  19.9 & 3.63 $\pm$0.02 $\pm$0.14 \\
 21.5-- & 25.1 &  23.2 & 2.48 $\pm$0.02 $\pm$0.10 \\
 25.1-- & 29.3 &  27.1 & 1.62 $\pm$0.01 $\pm$0.06 \\
 29.3-- & 34.1 &  31.6 & 1.09 $\pm$0.01 $\pm$0.04 \\
 34.1-- & 39.8 &  36.8 & 7.17 $\pm$0.08 $\pm$0.30 $\times 10^{-1}$\\
 39.8-- & 46.4 &  42.9 & 4.84 $\pm$0.06 $\pm$0.21 $\times 10^{-1}$\\
 46.4-- & 54.1 &  50.0 & 3.15 $\pm$0.05 $\pm$0.14 $\times 10^{-1}$\\
 54.1-- & 63.1 &  58.3 & 2.07 $\pm$0.03 $\pm$0.09 $\times 10^{-1}$\\
 63.1-- & 73.6 &  68.0 & 1.34 $\pm$0.03 $\pm$0.06 $\times 10^{-1}$\\
 73.6-- & 85.8 &  79.2 & 9.09 $\pm$0.19 $\pm$0.43 $\times 10^{-2}$\\
 85.8-- & 100. &  92.3 & 5.75 $\pm$0.14 $\pm$0.28 $\times 10^{-2}$\\
 100.-- & 126. &  112. & 3.43 $\pm$0.11 $\pm$0.18 $\times 10^{-2}$\\
 126.-- & 158. &  140. & 1.98 $\pm$0.07 $\pm$0.11 $\times 10^{-2}$\\
 158.-- & 200. &  177. & 1.00 $\pm$0.05 $\pm$0.06 $\times 10^{-2}$\\
 200.-- & 251. &  222. & 5.42 $\pm$0.31 $\pm$0.34 $\times 10^{-3}$\\
 251.-- & 316. &  281. & 2.46 $\pm$0.19 $\pm$0.17 $\times 10^{-3}$\\
 316.-- & 398. &  352. & 1.62 $\pm$0.14 $\pm$0.14 $\times 10^{-3}$\\
 398.-- & 541. &  463. & 7.47 $\pm$0.69 $\pm$0.78 $\times 10^{-4}$\\
      \hline \hline
    \end{tabular}
  \end{center}
\end{table}

\begin{table}
  \caption{Primary helium flux at the top of the atmosphere.}
  \scriptsize
  \label{tab:heflux}
  \vspace{0.5cm}
  \begin{center}
    \begin{tabular}{rlrl} \hline \hline
      \multicolumn{2}{c}{
	\begin{tabular}{@{}c@{}}       
	  Energy range\\ (GeV/n)
	\end{tabular}
      }
      & 
      \begin{tabular}{@{}c@{}}       
        $\overline{E_{\rm k}}$\\ (GeV/n)
      \end{tabular}
      &
      \begin{tabular}{@{}c@{}}       
	Flux$\pm\Delta F_{\rm sta}\pm\Delta F_{\rm sys}$\\
	(m$^{-2}$sr$^{-1}$s$^{-1}$(GeV/n$)^{-1}$)
      \end{tabular}
      \\
      \hline
 1.00-- & 1.17 &  1.08 & 5.22 $\pm$0.14 $\pm$0.36 $\times 10^{1}$\\
 1.17-- & 1.36 &  1.26 & 4.78 $\pm$0.13 $\pm$0.33 $\times 10^{1}$\\
 1.36-- & 1.58 &  1.47 & 4.02 $\pm$0.11 $\pm$0.28 $\times 10^{1}$\\
 1.58-- & 1.85 &  1.71 & 3.21 $\pm$0.09 $\pm$0.22 $\times 10^{1}$\\
 1.85-- & 2.15 &  2.00 & 2.62 $\pm$0.08 $\pm$0.18 $\times 10^{1}$\\
 2.15-- & 2.51 &  2.33 & 2.17 $\pm$0.06 $\pm$0.15 $\times 10^{1}$\\
 2.51-- & 2.93 &  2.71 & 1.81 $\pm$0.05 $\pm$0.12 $\times 10^{1}$\\
 2.93-- & 3.42 &  3.16 & 1.37 $\pm$0.04 $\pm$0.09 $\times 10^{1}$\\
 3.42-- & 3.98 &  3.69 & 9.77 $\pm$0.35 $\pm$0.67 \\
 3.98-- & 4.64 &  4.29 & 7.67 $\pm$0.28 $\pm$0.53 \\
 4.64-- & 5.41 &  4.98 & 5.71 $\pm$0.23 $\pm$0.39 \\
 5.41-- & 6.31 &  5.84 & 3.98 $\pm$0.06 $\pm$0.32 \\
 6.31-- & 7.36 &  6.80 & 2.83 $\pm$0.04 $\pm$0.22 \\
 7.36-- & 8.58 &  7.94 & 2.07 $\pm$0.03 $\pm$0.16 \\
 8.58-- & 10.0 &  9.24 & 1.48 $\pm$0.03 $\pm$0.12 \\
 10.0-- & 11.7 &  10.8 & 1.02 $\pm$0.02 $\pm$0.08 \\
 11.7-- & 13.6 &  12.6 & 6.76 $\pm$0.16 $\pm$0.54 $\times 10^{-1}$\\
 13.6-- & 15.8 &  14.7 & 4.71 $\pm$0.12 $\pm$0.38 $\times 10^{-1}$\\
 15.8-- & 18.5 &  17.0 & 3.27 $\pm$0.10 $\pm$0.26 $\times 10^{-1}$\\
 18.5-- & 21.5 &  19.9 & 2.13 $\pm$0.07 $\pm$0.17 $\times 10^{-1}$\\
 21.5-- & 25.1 &  23.2 & 1.46 $\pm$0.05 $\pm$0.12 $\times 10^{-1}$\\
 25.1-- & 29.3 &  27.1 & 9.67 $\pm$0.41 $\pm$0.79 $\times 10^{-2}$\\
 29.3-- & 34.1 &  31.4 & 5.89 $\pm$0.30 $\pm$0.48 $\times 10^{-2}$\\
 34.1-- & 39.8 &  36.7 & 4.23 $\pm$0.24 $\pm$0.35 $\times 10^{-2}$\\
 39.8-- & 46.4 &  42.9 & 2.85 $\pm$0.18 $\pm$0.24 $\times 10^{-2}$\\
 46.4-- & 54.1 &  49.9 & 1.84 $\pm$0.13 $\pm$0.15 $\times 10^{-2}$\\
 54.1-- & 73.6 &  62.5 & 9.40 $\pm$0.92 $\pm$0.90 $\times 10^{-3}$\\
 73.6-- & 100. &  86.1 & 4.14 $\pm$0.53 $\pm$0.41 $\times 10^{-3}$\\
 100.-- & 136. &  116. & 2.16 $\pm$0.33 $\pm$0.22 $\times 10^{-3}$\\
 136.-- & 251. &  175. & 5.53 $\pm$0.92 $\pm$0.64 $\times 10^{-4}$\\
      \hline \hline
    \end{tabular}
  \end{center}
\end{table}

\begin{table}
  \caption{Atmospheric muon flux at sea level.}
  \scriptsize
  \label{tab:muflux}
  \vspace{0.5cm}
  \begin{center}
    \begin{tabular}{rrrlrl}
      \hline
      \hline
      &	& \multicolumn{2}{c}{$\mu^+$} 
	& \multicolumn{2}{c}{$\mu^-$}\\
	\cline{3-6}
      \multicolumn{2}{l}{
	\begin{tabular}{@{}c@{}}       
	  Momentum\\ range\\ (GeV/$c$)
	\end{tabular}
      }
      & 
      \begin{tabular}{@{}c@{}}\\
	$\overline{P}$\\ (GeV/$c$)
      \end{tabular}
      &
      \begin{tabular}{@{}c@{}}\\
	Flux$\pm\Delta F_{\rm sta}\pm\Delta F_{\rm sys}$\\
	 (m$^{-2}$sr$^{-1}$s$^{-1}$(GeV/$c$)$^{-1}$)
      \end{tabular}
      & 
      \begin{tabular}{@{}c@{}}\\
	$\overline{P}$\\ (GeV/$c$)
      \end{tabular}
      &
      \begin{tabular}{@{}c@{}}\\
	Flux$\pm\Delta F_{\rm sta}\pm\Delta F_{\rm sys}$\\
	 (m$^{-2}$sr$^{-1}$s$^{-1}$(GeV/$c$)$^{-1}$)
      \end{tabular}\\
      \hline
0.576-- &0.621 & 0.599 & 1.34 $\pm$0.02 $\pm$0.03 $\times 10^{1}$ & 0.598 & 1.25 $\pm$0.02 $\pm$0.03 $\times 10^{1}$\\
0.621-- &0.669 & 0.645 & 1.33 $\pm$0.02 $\pm$0.03 $\times 10^{1}$ & 0.644 & 1.22 $\pm$0.02 $\pm$0.03 $\times 10^{1}$\\
0.669-- &0.720 & 0.694 & 1.32 $\pm$0.02 $\pm$0.03 $\times 10^{1}$ & 0.694 & 1.19 $\pm$0.02 $\pm$0.03 $\times 10^{1}$\\
0.720-- &0.776 & 0.748 & 1.27 $\pm$0.02 $\pm$0.03 $\times 10^{1}$ & 0.748 & 1.15 $\pm$0.02 $\pm$0.02 $\times 10^{1}$\\
0.776-- &0.836 & 0.806 & 1.26 $\pm$0.02 $\pm$0.03 $\times 10^{1}$ & 0.806 & 1.13 $\pm$0.01 $\pm$0.02 $\times 10^{1}$\\
0.836-- &0.901 & 0.869 & 1.23 $\pm$0.01 $\pm$0.03 $\times 10^{1}$ & 0.868 & 1.08 $\pm$0.01 $\pm$0.02 $\times 10^{1}$\\
0.901-- &0.970 & 0.934 & 1.21 $\pm$0.01 $\pm$0.03 $\times 10^{1}$ & 0.936 & 1.06 $\pm$0.01 $\pm$0.02 $\times 10^{1}$\\
0.970-- & 1.04 &  1.01 & 1.16 $\pm$0.01 $\pm$0.02 $\times 10^{1}$ &  1.01 & 1.00 $\pm$0.01 $\pm$0.02 $\times 10^{1}$\\
 1.04-- & 1.13 &  1.08 & 1.10 $\pm$0.01 $\pm$0.02 $\times 10^{1}$ &  1.08 & 0.96 $\pm$0.01 $\pm$0.02 $\times 10^{1}$\\
 1.13-- & 1.21 &  1.17 & 1.07 $\pm$0.01 $\pm$0.02 $\times 10^{1}$ &  1.17 & 0.91 $\pm$0.01 $\pm$0.02 $\times 10^{1}$\\
 1.21-- & 1.31 &  1.26 & 1.02 $\pm$0.01 $\pm$0.02 $\times 10^{1}$ &  1.26 & 0.88 $\pm$0.01 $\pm$0.02 $\times 10^{1}$\\
 1.31-- & 1.41 &  1.36 & 9.78 $\pm$0.10 $\pm$0.21  &  1.36 & 8.53 $\pm$0.10 $\pm$0.18 \\
 1.41-- & 1.52 &  1.46 & 9.38 $\pm$0.10 $\pm$0.20  &  1.46 & 8.01 $\pm$0.09 $\pm$0.17 \\
 1.52-- & 1.63 &  1.57 & 8.72 $\pm$0.09 $\pm$0.18  &  1.57 & 7.53 $\pm$0.08 $\pm$0.16 \\
 1.63-- & 1.76 &  1.70 & 8.59 $\pm$0.09 $\pm$0.18  &  1.70 & 7.22 $\pm$0.08 $\pm$0.15 \\
 1.76-- & 1.90 &  1.83 & 7.85 $\pm$0.08 $\pm$0.17  &  1.83 & 6.72 $\pm$0.07 $\pm$0.14 \\
 1.90-- & 2.04 &  1.97 & 7.41 $\pm$0.07 $\pm$0.16  &  1.97 & 6.28 $\pm$0.07 $\pm$0.13 \\
 2.04-- & 2.20 &  2.12 & 7.03 $\pm$0.07 $\pm$0.15  &  2.12 & 5.71 $\pm$0.06 $\pm$0.12 \\
 2.20-- & 2.37 &  2.28 & 6.38 $\pm$0.06 $\pm$0.13  &  2.29 & 5.36 $\pm$0.06 $\pm$0.11 \\
 2.37-- & 2.55 &  2.46 & 6.01 $\pm$0.06 $\pm$0.13  &  2.46 & 4.92 $\pm$0.05 $\pm$0.10 \\
 2.55-- & 2.75 &  2.65 & 5.45 $\pm$0.05 $\pm$0.13  &  2.65 & 4.62 $\pm$0.05 $\pm$0.10 \\
 2.75-- & 2.96 &  2.86 & 5.02 $\pm$0.05 $\pm$0.12  &  2.86 & 4.09 $\pm$0.05 $\pm$0.09 \\
 2.96-- & 3.19 &  3.08 & 4.62 $\pm$0.05 $\pm$0.11  &  3.08 & 3.69 $\pm$0.04 $\pm$0.08 \\
 3.19-- & 3.44 &  3.32 & 4.17 $\pm$0.04 $\pm$0.10  &  3.31 & 3.31 $\pm$0.04 $\pm$0.07 \\
 3.44-- & 3.71 &  3.57 & 3.79 $\pm$0.04 $\pm$0.09  &  3.57 & 3.06 $\pm$0.04 $\pm$0.06 \\
 3.71-- & 3.99 &  3.85 & 3.37 $\pm$0.04 $\pm$0.08  &  3.85 & 2.70 $\pm$0.03 $\pm$0.06 \\
 3.99-- & 4.30 &  4.14 & 3.02 $\pm$0.03 $\pm$0.07  &  4.14 & 2.37 $\pm$0.03 $\pm$0.05 \\
 4.30-- & 4.63 &  4.47 & 2.74 $\pm$0.03 $\pm$0.06  &  4.47 & 2.11 $\pm$0.03 $\pm$0.04 \\
 4.63-- & 4.99 &  4.81 & 2.41 $\pm$0.03 $\pm$0.06  &  4.81 & 1.87 $\pm$0.02 $\pm$0.04 \\
 4.99-- & 5.38 &  5.18 & 2.11 $\pm$0.02 $\pm$0.05  &  5.18 & 1.66 $\pm$0.02 $\pm$0.04 \\
 5.38-- & 5.79 &  5.58 & 1.87 $\pm$0.02 $\pm$0.04  &  5.58 & 1.46 $\pm$0.02 $\pm$0.03 \\
      \hline \hline
    \end{tabular}
  \end{center}
\end{table}

\begin{table}
  Table 3\\
  Atmospheric muon flux at sea level (continued).
  \scriptsize
  \vspace{0.5cm}
  \begin{center}
    \begin{tabular}{rrrlrl}
      \hline
      \hline
      & & \multicolumn{2}{c}{$\mu^+$} 
        & \multicolumn{2}{c}{$\mu^-$}\\
	\cline{3-6}
      \multicolumn{2}{l}{
	\begin{tabular}{@{}c@{}}       
	  Momentum\\ range\\ (GeV/$c$)
	\end{tabular}
      }
      & 
      \begin{tabular}{@{}c@{}}\\
	$\overline{P}$\\ (GeV/$c$)
      \end{tabular}
      &
      \begin{tabular}{@{}c@{}}\\
	Flux$\pm\Delta F_{\rm sta}\pm\Delta F_{\rm sys}$\\
	 (m$^{-2}$sr$^{-1}$s$^{-1}$(GeV/$c$)$^{-1}$)
      \end{tabular}
      & 
      \begin{tabular}{@{}c@{}}\\
	$\overline{P}$\\ (GeV/$c$)
      \end{tabular}
      &
      \begin{tabular}{@{}c@{}}\\
	Flux$\pm\Delta F_{\rm sta}\pm\Delta F_{\rm sys}$\\
	 (m$^{-2}$sr$^{-1}$s$^{-1}$(GeV/$c$)$^{-1}$)
      \end{tabular}\\
      \hline
 5.79-- & 6.24 &  6.01 & 1.63 $\pm$0.02 $\pm$0.04  &  6.01 & 1.24 $\pm$0.02 $\pm$0.03 \\
 6.24-- & 6.73 &  6.48 & 1.44 $\pm$0.02 $\pm$0.03  &  6.48 & 1.13 $\pm$0.02 $\pm$0.02 \\
 6.73-- & 7.25 &  6.98 & 1.21 $\pm$0.02 $\pm$0.03  &  6.98 & 0.96 $\pm$0.01 $\pm$0.02 \\
 7.25-- & 7.81 &  7.52 & 1.06 $\pm$0.01 $\pm$0.02  &  7.52 & 0.83 $\pm$0.01 $\pm$0.02 \\
 7.81-- & 8.41 &  8.10 & 9.11 $\pm$0.13 $\pm$0.21 $\times 10^{-1}$ &  8.10 & 6.93 $\pm$0.11 $\pm$0.15 $\times 10^{-1}$\\
 8.41-- & 9.06 &  8.72 & 8.07 $\pm$0.11 $\pm$0.19 $\times 10^{-1}$ &  8.72 & 6.08 $\pm$0.10 $\pm$0.13 $\times 10^{-1}$\\
 9.06-- & 9.76 &  9.40 & 7.06 $\pm$0.10 $\pm$0.16 $\times 10^{-1}$ &  9.40 & 5.15 $\pm$0.09 $\pm$0.11 $\times 10^{-1}$\\
 9.76-- & 10.5 &  10.1 & 5.67 $\pm$0.09 $\pm$0.13 $\times 10^{-1}$ &  10.1 & 4.54 $\pm$0.08 $\pm$0.10 $\times 10^{-1}$\\
 10.5-- & 11.3 &  10.9 & 4.90 $\pm$0.08 $\pm$0.11 $\times 10^{-1}$ &  10.9 & 3.85 $\pm$0.07 $\pm$0.08 $\times 10^{-1}$\\
 11.3-- & 12.2 &  11.8 & 4.38 $\pm$0.07 $\pm$0.10 $\times 10^{-1}$ &  11.8 & 3.24 $\pm$0.06 $\pm$0.07 $\times 10^{-1}$\\
 12.2-- & 13.2 &  12.7 & 3.58 $\pm$0.06 $\pm$0.08 $\times 10^{-1}$ &  12.7 & 2.74 $\pm$0.06 $\pm$0.06 $\times 10^{-1}$\\
 13.2-- & 14.2 &  13.6 & 3.07 $\pm$0.06 $\pm$0.07 $\times 10^{-1}$ &  13.7 & 2.31 $\pm$0.05 $\pm$0.05 $\times 10^{-1}$\\
 14.2-- & 15.3 &  14.7 & 2.49 $\pm$0.05 $\pm$0.06 $\times 10^{-1}$ &  14.7 & 1.97 $\pm$0.04 $\pm$0.04 $\times 10^{-1}$\\
 15.3-- & 16.4 &  15.8 & 2.13 $\pm$0.04 $\pm$0.05 $\times 10^{-1}$ &  15.8 & 1.68 $\pm$0.04 $\pm$0.04 $\times 10^{-1}$\\
 16.4-- & 17.7 &  17.1 & 1.76 $\pm$0.04 $\pm$0.04 $\times 10^{-1}$ &  17.0 & 1.40 $\pm$0.03 $\pm$0.03 $\times 10^{-1}$\\
 17.7-- & 19.1 &  18.4 & 1.49 $\pm$0.03 $\pm$0.03 $\times 10^{-1}$ &  18.4 & 1.15 $\pm$0.03 $\pm$0.02 $\times 10^{-1}$\\
 19.1-- & 20.6 &  19.8 & 1.22 $\pm$0.03 $\pm$0.03 $\times 10^{-1}$ &  19.8 & 0.96 $\pm$0.03 $\pm$0.02 $\times 10^{-1}$\\
 20.6-- & 23.9 &  22.1 & 9.57 $\pm$0.09 $\pm$0.22 $\times 10^{-2}$ &  22.1 & 7.36 $\pm$0.08 $\pm$0.15 $\times 10^{-2}$\\
 23.9-- & 27.7 &  25.6 & 6.69 $\pm$0.07 $\pm$0.16 $\times 10^{-2}$ &  25.6 & 5.04 $\pm$0.06 $\pm$0.11 $\times 10^{-2}$\\
 27.7-- & 32.1 &  29.7 & 4.52 $\pm$0.06 $\pm$0.10 $\times 10^{-2}$ &  29.8 & 3.48 $\pm$0.05 $\pm$0.07 $\times 10^{-2}$\\
 32.1-- & 37.3 &  34.5 & 2.93 $\pm$0.04 $\pm$0.07 $\times 10^{-2}$ &  34.5 & 2.26 $\pm$0.04 $\pm$0.05 $\times 10^{-2}$\\
 37.3-- & 43.3 &  40.1 & 2.01 $\pm$0.03 $\pm$0.05 $\times 10^{-2}$ &  40.1 & 1.55 $\pm$0.03 $\pm$0.03 $\times 10^{-2}$\\
 43.3-- & 50.2 &  46.5 & 1.31 $\pm$0.02 $\pm$0.03 $\times 10^{-2}$ &  46.5 & 1.04 $\pm$0.02 $\pm$0.02 $\times 10^{-2}$\\
 50.2-- & 58.3 &  54.0 & 8.76 $\pm$0.18 $\pm$0.22 $\times 10^{-3}$ &  54.0 & 6.91 $\pm$0.16 $\pm$0.16 $\times 10^{-3}$\\
 58.3-- & 67.7 &  62.7 & 5.72 $\pm$0.14 $\pm$0.15 $\times 10^{-3}$ &  62.7 & 4.54 $\pm$0.12 $\pm$0.11 $\times 10^{-3}$\\
 67.7-- & 78.5 &  72.8 & 4.01 $\pm$0.11 $\pm$0.11 $\times 10^{-3}$ &  72.8 & 2.88 $\pm$0.09 $\pm$0.08 $\times 10^{-3}$\\
 78.5-- & 91.1 &  84.4 & 2.55 $\pm$0.08 $\pm$0.08 $\times 10^{-3}$ &  84.5 & 1.93 $\pm$0.07 $\pm$0.06 $\times 10^{-3}$\\
 91.1-- & 106. &  98.0 & 1.55 $\pm$0.06 $\pm$0.05 $\times 10^{-3}$ &  97.7 & 1.17 $\pm$0.05 $\pm$0.04 $\times 10^{-3}$\\
 106.-- & 132. &  118. & 9.43 $\pm$0.43 $\pm$0.40 $\times 10^{-4}$ &  117. & 6.95 $\pm$0.37 $\pm$0.28 $\times 10^{-4}$\\
 132.-- & 165. &  147. & 4.26 $\pm$0.26 $\pm$0.22 $\times 10^{-4}$ &  146. & 3.65 $\pm$0.24 $\pm$0.18 $\times 10^{-4}$\\
 165.-- & 207. &  184. & 2.51 $\pm$0.18 $\pm$0.15 $\times 10^{-4}$ &  185. & 1.47 $\pm$0.14 $\pm$0.09 $\times 10^{-4}$\\
 207.-- & 258. &  229. & 1.18 $\pm$0.11 $\pm$0.09 $\times 10^{-4}$ &  228. & 0.75 $\pm$0.09 $\pm$0.06 $\times 10^{-4}$\\
 258.-- & 323. &  286. & 5.58 $\pm$0.68 $\pm$0.52 $\times 10^{-5}$ &  286. & 4.13 $\pm$0.58 $\pm$0.39 $\times 10^{-5}$\\
 323.-- & 404. &  359. & 2.56 $\pm$0.41 $\pm$0.30 $\times 10^{-5}$ &  355. & 2.45 $\pm$0.40 $\pm$0.28 $\times 10^{-5}$\\
      \hline \hline
    \end{tabular}
  \end{center}
\end{table}

\end{document}